\documentstyle[prc,aps,twocolumn,ifthen,epsf,epsfig]{revtex}

\newcommand{\nc}{\newcommand}
\nc{\beq}{\begin{equation}}
\nc{\eeq}{\end{equation}}

\newcommand{\prevc}[3]{Phys. Rev. {\bf C#1}, #3 (#2)}
\newcommand{\prevd}[3]{Phys. Rev. {\bf D#1}, #3 (#2)}
\newcommand{\prevl}[3]{Phys. Rev. Lett.\ {\bf #1}, #3 (#2)}

\newcommand{\plb}[3]{Phys. Lett. {\bf B#1}, #3 (#2)}
\newcommand{\npa}[3]{Nucl. Phys. {\bf A#1}, #3 (#2)}

\begin{document}
\twocolumn[\hsize\textwidth\columnwidth\hsize\csname @twocolumnfalse\endcsname
\title{$v_4$: A small, but sensitive observable for heavy ion collisions}
\author{Peter F. Kolb}
\address{Department of Physics and Astronomy, 
             SUNY, Stony Brook, New York, 11794-3800, USA}
\date{September 23, 2003}

\maketitle


\begin{abstract}
Higher order Fourier coefficients of the azimuthally dependent 
single particle spectra resulting from noncentral heavy ion collisions are investigated. 
For intermediate to large transverse momenta, these anisotropies are expected to become
as large as 5 \%, and should be clearly measurable.
The physics content of these observables is discussed from two different 
extreme but complementary viewpoints, hydrodynamics and the geometric 
limit with extreme energy loss.
\end{abstract}
\vspace*{0.2in}
]


Collisions of large nuclei at nonzero impact parameter exhibit the special feature 
of a strongly deformed overlap region. The subsequent dynamical evolution of the
collision zone converts the initial spatial eccentricity into an azimuthal anisotropy 
of the final state observables.
In particular anisotropies in momentum space~\cite{Ollitrault92} 
have been analyzed experimentally in great detail in recent years, 
as they have been shown to be generated during the earliest and hottest stages of 
the reaction~\cite{Sorge97}. 
As momentum anisotropies are generated, the eccentricity
in coordinate space is smoothed out, bringing further generation of momentum 
anisotropies to a stall, a process that occurs within about the first 5 fm/$c$ of the total
15 fm/$c$ lifetime of the fireball~\cite{Sorge97,ZGK99,KSH00}.
In Au+Au collisions at RHIC (the Relativistic Heavy Ion Collider 
at Brookhaven National Laboratory), those anisotropies have been 
found~\cite{STARfirstv2} to be as large as predicted by 
hydrodynamic calculations that assume a short equilibration time 
$\tau_{\rm equ} \leq 1$ fm/$c$~\cite{KSH00,KHHH01}. 
Further theoretical investigation has revealed the sensitivity of anisotropic flow 
on the equation of state of the expanding medium~\cite{TLS01,HKHRV01,KH02},
indicating the necessity of a hard phase at high temperatures with 
a soft transition region of width $\Delta e \sim 1$~GeV/fm$^3$ to 
lower temperatures in order to describe the data in more detail.
For these reasons anisotropic flow became one of the most attractive tools 
to study the nuclear equation of state of the matter created in the collision 
by utilizing the most abundant particles, 
alternatively to study statistically disadvantaged rare probes with which one tries to
infer the properties of the bulk by modeling their mutual interaction with the medium 
they traverse.
Other predictions from hydrodynamic concepts such as centrality and 
mass dependence of  momentum anisotropies~\cite{HKHRV01}
have subsequently been confirmed to be in qualitative and quantitative agreement 
with experiments~\cite{STARv2s130identifieds,PHENIXv2s130centralities}, as long as the applicability of hydrodynamics
is not overstretched, i.e. for impact parameters not larger than $b \sim 7$~fm and transverse 
momenta not to exceed $p_T \sim 1.5$~GeV. 
Beyond these values clear deviations from the
hydrodynamic predictions start to occur. 
This is expected as the smaller systems do not sufficiently equilibrate and 
the most rapid particles escape the fireball without fully participating in 
the collective dynamics.
%

\begin{figure}[h,t,b,p]
 \hspace*{-.40cm}
  \epsfig{file=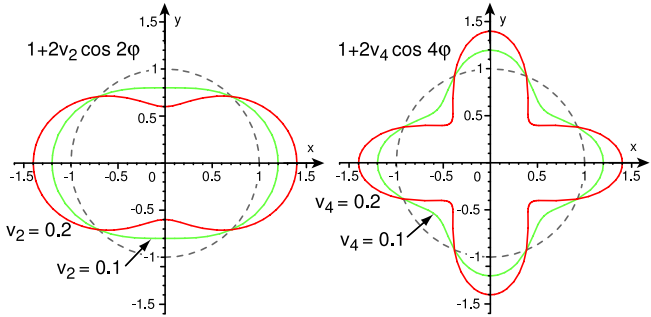,width=8.5cm}
 \vspace*{3mm}
 \caption{(Color online)
	Polar plot of the azimuthal distribution of different Fourier expansions. 
	Left, distortion of the unit circle through an elliptic component $v_2$,
	 shown for 10\% (green) and 20\% (red).
	Right, fourth component $v_4$ for the same values.          }
 \label{fig:polarplots}
 \end{figure}

%
To date, anisotropies at midrapidity are generally only characterized in terms of the 
second coefficient $v_2$ of the Fourier decomposition of the azimuthally sensitive 
momentum spectrum. 
With the $z$ direction given by the beam axis and the
$x$ direction defined by the direction of the impact parameter (and $y$ perpendicular
to both),  the general expression for the Fourier decomposition with respect to the 
azimuthal angle $\varphi = \arctan (y/x)$ is
\beq
\label{Fourierexpansion}
\frac{dN}{dp_T dy d\varphi}
=
\frac{1}{2 \pi}
\frac{dN}{dp_T dy}
\left( 1 + \!\! \sum_{n, \,{\rm even}} \! \! 2 \, v_n(p_T) \cos (n \varphi) \right),
\eeq
where no sine terms appear due to the symmetry with respect to the reaction plane, 
and all odd harmonics vanish at midrapidity due to the 
symmetry $\varphi \leftrightarrow \varphi+\pi$. 
(Note that the latter is not true for noncentral $d$-Au collisions, 
where $v_1$, $v_3$,... should be useful quantities to classify and study
the measured spectra -- even at midrapidity).
In the following we will elaborate on the centrality and momentum dependence of 
higher anisotropy coefficients, which turn out to exhibit interesting features that, 
with the currently  available data from RHIC's run in 2002, might already be accessible.
Generally,  higher momentum anisotropies are expected to be small, and in fact no 
experimental data have been published until this day. 
An early hydrodynamic calculation~\cite{KSH99}  predicted the momentum integrated 
value of $v_4$ not to exceed 0.5 \% at SPS energies even in the most peripheral 
collisions, and shows  only a slight increase with beam energy up to RHIC energies. 
In another study Teaney and Shuryak~\cite{TS99} calculated $\alpha_4$, 
which is the fourth coefficient of the momentum integrated differential particle spectra 
weighted by an additional momentum squared 
(which was applied to enhance the effects of transverse flow). 
This quantity is thus strongly biased to larger $p_T$. 
From their hydrodynamic calculation they obtain $\alpha_4 \sim 1 \%$; 
clearly the momentum integrated value of $v_4$ is still smaller.
Today, experiments at RHIC achieve sufficient statistics to measure the
elliptic flow coefficient $v_2$ differentially in momentum out to large 
transverse momenta up to 10 GeV~\cite{STARv2highpt}. 
In midcentral collisions this coefficient is found to saturate
starting at $p_T\sim 2$~GeV at a large value of 20 -- 25\%, indicating that three times 
as many particles are emitted into the reaction plane than out of the reaction plane. 
Such a large value also means  that the first order deformation of Eq. (\ref{Fourierexpansion}) 
is not elliptic anymore, but that a polar plot of the azimuthal distribution resembles 
more a peanut than an elliptical shape, aligned with its longer axis into
the reaction plane, see Fig. \ref{fig:polarplots}.
Without any higher order coefficients, the elliptic deformation starts
to develop such a waist when $v_2$ becomes larger than 10~\%,
which can be shown explicitly by discussing the term $1+\cos 2 \varphi$ in 
Cartesian coordinates $(x,y)$ around $x=0$.
%

%
%

%
This raises the interesting question, whether at high $p_T$  Fourier coefficients of
higher order might become sufficiently large to restore the elliptically deformed shape of 
the particle distribution.
To restore the elliptic shape as much as possible only through introduction of
a fourth anisotropy coefficient, $v_4 \sim \frac{1}{34}(10\, v_2-1)$ is required, as
is derived straightforwardly from Eq. (\ref{Fourierexpansion}).\footnote{
The definition of restored ellipticity is ambiguous. Here we chose to find
the value of $v_4$ where the second derivative of $y(x)$ vanishes at $x=0$, 
thus giving rise to a very smooth transition across $\varphi=\pi/2$.
} 
This, in consideration of the large experimental values for $v_2$, might indicate 
that $v_4$ could reach values around 3--5~\%.

\begin{figure}[h,t,b,p]
 \hspace*{.5cm}
  \epsfig{file=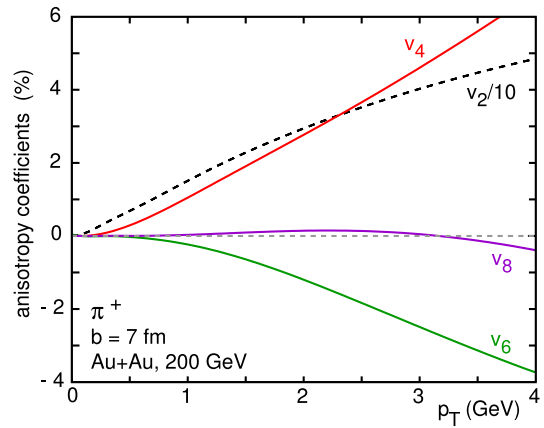,width=7cm}
 \vspace*{3mm}
 \caption{(Color online)
	Fourier coefficients of the pion spectra resulting from a hydrodynamic calculation
	to describe particle spectra at $\sqrt{s_{\rm NN}}=200$~GeV at impact parameter 
	$b=7$~fm.
	$v_2$ is scaled by a factor 0.1.       }
 \label{fig:hydroanisotropies}
 \end{figure}

%
To investigate higher flow coefficients in more detail,  we first revisit 
hydrodynamic calculations, although they find their limitations 
for pions of transverse momenta beyond 2 GeV, and for heavier baryons at momenta 
3--4~GeV (for a recent review see Ref. \cite{KH03}).
Later on we will discuss a complementary model in terms of a simple picture for the 
limit of very large transverse momentum.
The results presented in the following were achieved by a more modern hydrodynamic
calculation than in Ref.~\cite{KSH99}, as it becomes important at RHIC energies to 
correctly account for features of chemical non-equilibrium during the hadronic
evolution of the system~\cite{Rapp02,KR02}, a feature that significantly influences the 
relation of energy-density and temperature~\cite{Teaney02,HT02}.
At temperatures beyond $T_{\rm crit}=165$~MeV the underlying equation of state
turns into a hard ideal gas equation of state via a strong first order phase transition in 
order to mimic the transition from a gas of interacting resonances to a hard
plasma phase.
The initial conditions of this calculation were determined to fit single particle spectra measured 
in Au+Au collisions at $\sqrt{s_{\rm NN}}= 200$~GeV. For further details, 
see Ref.~\cite{KR02}.
Figure~\ref{fig:hydroanisotropies} shows the momentum dependence of 
the Fourier coefficients of pion spectra up to order 8. 
``Elliptic flow,'' $v_2$ dominates the emission anisotropy at all momenta. 
(To use the same vertical scale for all coefficients, the second harmonic $v_2$
was divided by 10.) 
As anticipated, $v_4$ increases with momentum, reaching a value of 
about 2.8 \% at $p_T=2$~GeV (thus not quite enough to restore the particle distribution 
to an elliptic shape, as $v_2(p_T=2\,{\rm GeV})\approx 29 \%$).
Also $v_6$ still has a sizable {\it negative} value of about 1.2\%.  
The mass dependence of higher flow harmonics does not bring any surprises. 
Due to their larger masses, heavy particles pick up larger transverse momenta when they  
participate in the established collective flow. 
As for the spectra and elliptic flow, also the higher anisotropies are shifted 
out toward larger transverse momenta. 
Thus at a given transverse momentum below 2 GeV, the anisotropies of 
heavy resonances are smaller than of pions. 
Comparing hydrodynamic calculations to experimental results, 
it appears however that deviations from hydrodynamic behavior 
occur later for heavy particles, and therefore  anisotropies of baryons 
can eventually surpass anisotropies of light  mesons~\cite{v2mesonbaryon}. 
From these calculations one can expect a fourth flow component of (anti-)protons of 
about 7 \% at  4 GeV transverse momentum, a sizable quantity that should be
measurable.

%
%

\begin{figure}[h,t,b,p]
 \hspace*{.2cm}
 \epsfig{file=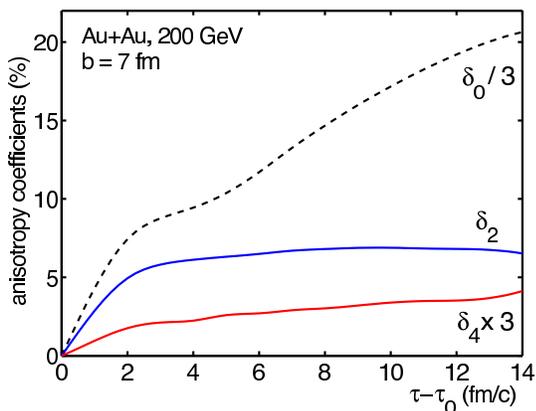,width=7cm}
 \caption{(Color online)
	Time evolution of the anisotropy coefficients in the transverse velocity field
	as given by the hydrodynamic calculation. $\delta_0$ 
	characterizes the radial flow,
	$\delta_2$ the elliptic deformation and $\delta_4$ 
	the quadrupole deformation.}
 \label{fig:anisoevolution}
 \end{figure}

Surprisingly it turns out that $v_4$ is highly sensitive to the initial conditions of the 
calculations: initializing the calculation with a prehydrodynamic, isotropic radial flow field
as in Ref.~\cite{KR02} does not change the magnitude of the flow anisotropies by a large
amount. 
However we find that it completely changes the sign of $v_4$, while the other
coefficients remain largely the same. Clearly $v_4$ at low and intermediate $p_T$ has 
therefore a strong potential to constrain model calculations and carries valuable 
information on the dynamical evolution of the system.
To study how the higher flow anisotropies develop in the course of the hydrodynamic 
evolution we calculate moments of the transverse flow field $(v_x,\, v_y)$ through
\beq
\label{flowansio}
\delta_n (\tau)= \frac{\int dx dy \, e(x,y; \tau)\, \gamma \, v_T \cos(n \varphi_v)}
                                    {\int dx dy \, e(x,y; \tau)\, \gamma \, }\,,
\eeq
where $e(x,y; \tau)$ is the energy density in the transverse plane at time $\tau$
and $v_T=(v_x^2+v_y^2)^{1/2}$ is the transverse flow velocity at a given point $(x,y)$,
$\gamma = (1-v_T^2)^{-1/2}$, and $\varphi_v = \arctan (v_y/v_x)$ is the angle of the 
local flow velocity with respect to the reaction plane. 
(Note that for $n=0$ this reduces to the definition of the mean 
radial velocity used in earlier studies \cite{KSH00}, but for $n=2$ it is slightly different from 
the momentum anisotropy $\epsilon_p$ which is defined in terms of the difference in the
diagonal elements of the energy momentum tensor.)
Figure \ref{fig:anisoevolution}  displays the time evolution of the flow coefficients and 
clearly shows that radial flow $\delta_0$
increases throughout the lifetime of the  system (which is about 15 fm/$c$), 
with a somewhat reduced acceleration while most of  the system is 
in the mixed phase at around 4 fm/$c$.
Quite oppositely anisotropy components quickly saturate and remain constant, 
as the initial geometric deformation of the source is lost. 
About 2/3 of the final value of $\delta_4$ is generated during the earliest stages
of the evolution, where the system is governed by the hard equation of state of the
quark gluon plasma.
$\delta_6$ quickly achieves small negative values of the order of -0.2 \% 
(as expected from the negative value of $v_6$). Due to its small values, it is
subject to numerical fluctuations and therefore not shown. 
It is expected that the flow coefficients are monotonically related to the 
anisotropy coefficients in the final particle spectra. Details of this relation 
are subject to future study.

%
%

\begin{figure}[h,t,b,p]
 \hspace*{.5cm}
  \epsfig{file=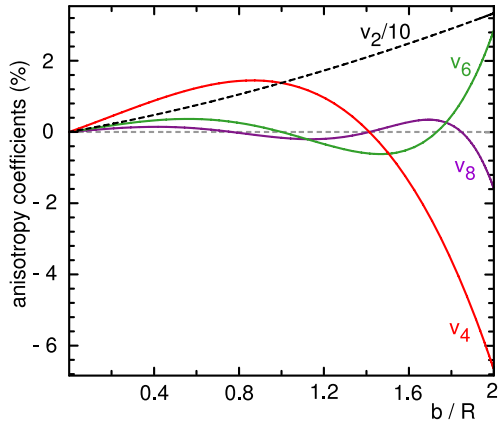,width=6.5cm}
 \vspace*{3mm}
 \caption{(Color online)
	Centrality dependence of the anisotropy coefficients in the limit of 
	pure surface emission from the deformed overlap region 
                       to describe the saturation value of the anisotropies 
                       at large transverse momenta.
	$v_2$ is scaled by a factor of 0.1.}
 \label{fig:geometryanisotropies}
 \end{figure}

%
In a model complementary to the hydrodynamic description which is constrained
to the low momentum, thermalized part of the spectrum, one can study 
anisotropies that result from opacity effects that the  deformed initial geometry
exerts on particles of high momenta \cite{GVWZ03,HKH02,Shuryak02}.
Assuming extreme opacity, or extreme jet quenching, particles of large transverse 
momenta from any part of the surface can only be emitted within $\pm 90^\circ$ of the
normal vector to the surface, which for this exploratory study is assumed to be given
by the overlap of two hard spheres (radius $R$).\footnote{
Relaxing this geometrical constraint by adopting a finite diffuseness of 
the nuclei reduces the resulting asymmetries, far below 
the experimental results, and novel effects are required to explain those 
large limiting values \cite{CS03,v2coalescence}.
}
In this picture, the $n$th Fourier coefficient of the resulting expansion is found to be 
(for $n$ even)
\beq
\label{vnblacklense}
v_n=(-1)^{n/2} \frac{1}{1-n^2} \frac{\sin (n \alpha)}{n \alpha}\,,
\eeq
where $\alpha= \arccos (b/2R)$ characterizes the centrality in terms of the
opening angle between the line through the center of one nucleus to the 
intersection point of the two nuclear spheres and the reaction plane. 
Note that as $b$ increases from 0 to $2R$, $\alpha$ decreases from $\pi/2$ to 0.
For $n=2$, $v_2=\sin(2\alpha)/(6 \alpha)$,
a formula first given by Voloshin~\cite{Voloshin03}.
For $n>2$, Eq. \ref{vnblacklense} becomes significantly more interesting 
(unfortunately, however, the size of the expected signal drops with $n^{-3}$).
Note first the oscillating sign of the prefactor which will manifest
itself in an oscillation of the coefficients at  large $b$ (small $\alpha$). 
Second, one notices that for larger $n$ the sine function starts changing
sign upon scanning impact parameters ($b\leftrightarrow \alpha$), 
whereas for $n=2$, the Fourier coefficient
increases linearly with impact parameter from 0 to 1/3, and is thus always
positive. 
In particular for $n=4$ one observes a transition from positive to negative 
values at $b = \sqrt{2} R$, reaching the limiting value 
$v_4= - 1/15 \approx - 6.67 \, \%$ for $b \rightarrow 2R$, 
again a rather large number. 
The local maximum of $v_4$ is at $b \approx 0.865 R$ with $v_4 \approx 1.45~\%$.
The centrality dependences of the second to eighth  Fourier coefficients are shown
in Fig. \ref{fig:geometryanisotropies} (as before $v_2$ is divided by a factor of 10).

Clearly, the expected values for higher expansion coefficients from this calculation are
small. 
Applying a more realistic model does, however, not necessarily lead to a further reduction
of the signal, but might in contrast increase, similarly as observed for calculations of
elliptic flow \cite{CS03}. 
This is also suggested by the experimental data on $v_2$ which seem to exceed 
the limits given even by the most optimistic assumptions for anisotropies from geometry
and energy loss.
In any case, such a distinctive signal as a sign change upon variation of centrality should 
be preserved when model calculations get refined and should be accessible within 
the statistics reached by today's experiments.
%

%
%

%
%
As an aside we would like to discuss a new but naive experimental way of measuring 
higher order flow coefficients, in particular $v_4$. 
Apart from the ``straightforward'' method to determine each event's reaction plane and
averaging all particles' $\cos (4 \varphi)$ with respect to the reaction plane
and  the more recently developed cumulant technique~\cite{BDO01}, 
simply measuring the particle spectra with respect to the reaction plane might be 
sufficient (theoretically). 
Having the azimuthally sensitive particle spectra given, Eq. (\ref{Fourierexpansion})
shows directly that [abbreviating $dN/p_Tdp_Tdyd\varphi \, (p_T,\,\varphi)= n(\varphi)$]
\begin{eqnarray}
\label{v4fromspectra}
&&n(0)+n(\pi/2)-2n(\pi/4) = \hspace*{3cm} \ \nonumber \\ 
&&\hspace*{.5cm}=\frac{1}{2\pi} \frac{dN}{p_Tdp_T} \left[ 8 \, v_4(p_T) + 8 \, v_{12}(p_T) +... \right]\,.
\end{eqnarray}
From the experience gathered so far we expect $v_{12}$ to be negligibly
small compared to $v_4$. 
Finite opening angles for the azimuthally sensitive distribution will reduce the 
signal again, however, the $v_4$ is in this approach enhanced by a factor 
of 8, which might still render it measurable. 
We propose this method thus as an interesting independent approach 
to measure $v_4$ (but also $v_2$, by simply subtracting spectra in and out of the
reaction plane).
%

%
In summary, we have shown that $v_4(p_T)$ might achieve significantly large values at 
intermediate and large transverse momenta in noncentral heavy ion collisions and contains
important physics of heavy ion collisions, a fact largely overlooked in the past. 
Numerical estimates for its size were given both in the limit of full thermalization 
with a subsequent hydrodynamic expansion as well as in the geometrical limit of 
extreme ``jet quenching.'' 
For the latter case a strong centrality dependence and a change of the sign 
of the signal for higher  anisotropies was found.
From both approximations we found that $v_4$ might reach values of the order of 5\%,
clearly observable with todays equipment and statistics.
$v_4$ has been found to be highly dependent on the initial configuration of the 
system and its evolution and is thus an important new tool to constrain model 
calculations and analyze details of the system's history.
Consequently it deserves to be studied in great detail both  experimentally as well as 
theoretically.

{\bf Acknowledgments:}
Discussions with 
J. Casalderrey-Solana,
U. Heinz, 
B. Jacak,
J. Jia,
A. Poskanzer,
E. Shuryak, 
and
R. Snellings
are greatly appreciated.
This work was supported by the U.S. Department of Energy under
Grant No. DE-FG02-88ER40388 and by the Alexander von Humboldt Foundation.\\[-4mm]

 \end{document}